\documentstyle[12pt,cite]{article}

\def\bea{\begin{eqnarray}}
\def\eea{\end{eqnarray}}

\def\a{\alpha}
\def\b{\beta}

\def\e{\epsilon}

\def\g{\gamma}

\def\k{\kappa}

\def\m{\mu}
\def\n{\nu}

\def\t{\tau}

\def\D{\Delta}

\def\L{\Lambda}

\def\ve{\varepsilon}

\def\cl{{\cal L}}

\def\co{{\cal O}}

\def\10{{\bf 10}}
\def\hh{{\bf 5}}
\def\5b{{\bf 5^*}}
\def\1{{\bf 1}}

\def\bo{{\raise.15ex\hbox{\large$\Box$}}}               
\def\pr{\prod}                                          
\def\face{{\raise.2ex\hbox{$\displaystyle \bigodot$}\mskip-2.2mu \llap {$\ddot
        \smile$}}}                                      

\def\leftrightarrowfill{$\mathsurround=0pt \mathord\leftarrow \mkern-6mu
        \cleaders\hbox{$\mkern-2mu \mathord- \mkern-2mu$}\hfill
        \mkern-6mu \mathord\rightarrow$}       
\def\dvec#1{\vbox{\ialign{##\crcr
        \leftrightarrowfill\crcr\noalign{\kern-1pt\nointerlineskip}
        $\hfil\displaystyle{#1}\hfil$\crcr}}}           

\def\beq{\begin{equation}}
\def\eeq{\end{equation}}

\def\pl#1#2#3{Phys.~Lett.~{\bf B {#1}} (19{#2}) #3}
\def\np#1#2#3{Nucl.~Phys.~{\bf B {#1}} (19{#2}) #3}
\def\prl#1#2#3{Phys.~Rev.~Lett.~{\bf #1} (19{#2}) #3}
\def\pr#1#2#3{Phys.~Rev.~{\bf D {#1}} (19{#2}) #3}
\def\zp#1#2#3{Z.~Phys.~{\bf C {#1}} (19{#2}) #3}

\def\ptp#1#2#3{Progr.~Theor.~Phys.~{\bf {#1}} (19{#2}) #3}

\def\nc#1#2#3{Nuovo Cim.~{\bf {#1}} (19{#2}) #3}

\catcode`@=11
\def\@citex[#1]#2{\if@filesw\immediate\write\@auxout{\string\citation{#2}}\fi
  \def\@citea{}\@cite{\@for\@citeb:=#2\do
    {\@citea\def\@citea{,\penalty\@m}\@ifundefined
      {b@\@citeb}{{\bf ?}\@warning
       {Citation `\@citeb' on page \thepage \space undefined}}%
\hbox{\csname b@\@citeb\endcsname}}}{#1}}

\def\citer{\@ifnextchar [{\@tempswatrue\@citexr}{\@tempswafalse\@citexr[]}}
\def\@citexr[#1]#2{\if@filesw\immediate\write\@auxout{\string\citation{#2}}\fi
  \def\@citea{}\@cite{\@for\@citeb:=#2\do
    {\@citea\def\@citea{--\penalty\@m}\@ifundefined
       {b@\@citeb}{{\bf ?}\@warning
       {Citation `\@citeb' on page \thepage \space undefined}}%
\hbox{\csname b@\@citeb\endcsname}}}{#1}}
\catcode`@=12

\begin{document}
\date{\mbox{ }}

\title{ 
{\normalsize     
DESY 98-155\hfill\mbox{}\\
October 1998\hfill\mbox{}\\}
\vspace{2cm}
\bf QUARK LEPTON MASS HIERARCHIES\\ AND THE BARYON ASYMMETRY\\[8mm]}
%
\author{W.~Buchm\"uller$^a$, T.~Yanagida$^{a,b}$\\
{\it a Deutsches Elektronen-Synchrotron DESY, Hamburg, Germany}\\
{\it b Department of Physics, University of Tokyo, Tokyo, Japan}}
\maketitle

\thispagestyle{empty}

\vspace{1cm}
\begin{abstract}
\noindent
The mass hierarchies of quarks and charged leptons as well as a
large $\n_\m$-$\n_\t$ mixing angle are naturally explained by the
Frogatt-Nielsen mechanism with a nonparallel family structure of chiral 
charges. We extend this mechanism to right-handed neutrinos. Their 
out-of-equilibrium decay generates a cosmological baryon asymmetry whose 
size is quantized in powers of the hierarchy parameter $\e^2$.
For the simplest hierarchy pattern the neutrino mass $\bar{m}_\n=
(m_{\n_\m}m_{\n_\t})^{1/2} \sim 10^{-2}$~eV, which is inferred from present
indications for neutrino oscillations, implies a baryon asymmetry 
$n_B/s \sim 10^{-10}$. The corresponding baryogenesis temperature is 
$T_B \sim 10^{10}$~GeV.
\end{abstract}

\newpage

In the standard model left- and right-handed quarks and leptons can be grouped 
into the SU(5) multiplets $\10 = (q_L,u_R^c,e_R^c)$, $\5b = (d_R^c,l_L)$
and $\1 = \n_R^c$, such that the Yukawa interactions with the Higgs fields
take the form,
\beq\label{yuk}
\cl = h_{ij}^{(u)}\ \10_i\10_j H_1(\hh) +  h_{ij}^{(d)}\ \5b_i\10_j H_2(\5b)
      + h_{ij}^{(\n)}\ \5b_i \1_j H_1(\hh) + h_{ij}^{(s)}\ \1_i\1_j S(\1)\;.
\eeq
Here $i,j=1\ldots 3$ are generation indices. The expectation values of the 
Higgs multiplets $H_1$ and $H_2$ generate ordinary Dirac masses of quarks and 
leptons, whereas the expectation value of the singlet Higgs field $S$ generates
the Majorana mass matrix of the right handed neutrinos. 

The masses of up-quarks, down-quarks and charged leptons approximately 
satisfy the following mass relations,
\beq
m_t : m_c : m_u \simeq 1 : \e^2 : \e^4\;,
\eeq
\beq\label{5hier}
m_b : m_s : m_d \simeq m_\t : m_\m : m_e \simeq 1 : \e : \e^3\;,
\eeq
where, for masses defined  at the unification scale $\L_{GUT}$, 
$\e^2 \sim 1/300$. According to the Frogatt-Nielsen mechanism \cite{fro79}
these mass 
hierarchies are related to a spontaneously broken symmetry. Each SU(5) 
multiplet of fermions $\psi_i$ carries a corresponding charge $Q_i$, 
and the Yukawa couplings then scale like
\beq
h_{ij} \propto \e^{Q_i + Q_j}\;.
\eeq
It is usually assumed that the mass hierarchy is generated by the expectation 
value of a singlet field $\Phi$ with charge $Q_{\Phi}=-1$ via a
nonrenormalizable interaction with a scale $\L > \L_{GUT}$, i.e.
$\e = \langle \Phi\rangle/\L$.    

\begin{table}[b]
\begin{center}
\begin{tabular}{c|ccccccccc}\hline \hline
$\psi_i$ & $\10_3$ & $\10_2$ & $\10_1$ & $\5b_3$ & $\5b_2$ & $\5b_1$ &
$\1_3$ & $\1_2$ & $\1_1$ \\ \hline
$Q_i$ & 0 & 1 & 2 & a & a & a+1 & b & c & d \\ \hline\hline
\end{tabular}
\medskip
\caption[dum]{\it Chiral charges for the fermions of the standard model; 
a=0 or 1, and 
\mbox{\phantom{Table 1:}$\hspace{5mm} 0\leq b \leq c \leq d$}.}
\end{center}
\end{table}

The observed mass hierarchy of up-quarks and the large t-quark mass yield
for the $\10$-plets uniquely the charges listed in table~1. Here we have
assumed that Yukawa couplings are not larger than $\co(1)$. From the
mass hierarchy of the charged leptons one infers that the second and third 
generation $\5b$-plet have the same charge (a), which differs by one unit 
from the charge of the first generation $\5b$-plet. The value of the b-quark 
mass allows a=0 or a=1 \cite{sat98}. Such a nonparallel family structure
has previously been identified in a detailed study of U(1) generation
symmetries \cite{bij87}. Note, that the U(1) may be replaced by an anomaly-free
discrete $Z_5$ symmetry with the charge assignment in table~1.

The difference between the observed down-quark mass hierarchy and the 
charged lepton mass hierarchy can be
accounted for by introducing an additional ${\bf 45}$-plet of Higgs fields 
\cite{geo79}. The following discussion does not depend on this and could also 
be carried out directly for the quark and lepton multiplets of the standard 
model gauge group. The chiral charges of the lepton doublets would then
correspond to the charges of the $\5b$-plets given in table~1.

Masses for the standard model neutrinos are generated by the unique
dimension-5 operator
\beq
\cl_{\D L=2} = f_{ij}\ \5b_i \5b_j H_1(\hh) H_1(\hh)\;,
\eeq
which is induced by the exchange of heavy particles in generic seesaw models 
\cite{yan79}. The chiral charges of the $\5b$ plets lead to the neutrino mass
matrix ($v=\langle H_1\rangle$),
\beq\label{matrix}
m_{\n_{ij}}=f_{ij}v^2 \propto \left(\begin{array}{ccc}
    \e^2  & \e  & \e \\[1ex]
    \e  &  1  & 1 \\[1ex]
    \e  &  1  & 1 
    \end{array}\right)\;,
\eeq
where we have only listed the order in $\e$ for the different matrix elements.
Diagonalization clearly yields a large $\n_\m$-$\n_\t$ mixing angle,
which is needed to explain the atmospheric neutrino deficit by $\n_\m$-$\n_\t$
oscillations \cite{atm98}. As described above, this is a direct consequence 
of the mass hierarchy of the charged leptons which leads to the charge 
assignment given in table~1. The masses of the two eigenstates 
$\n_\m$ and $\n_\t$ depend on factors of order one, which are omitted 
in (\ref{matrix}), and may easily differ by an order of magnitude 
\cite{ram98}. They can therefore be consistent with the mass differences 
$\D m^2_{\n_e \n_\m}\simeq 4\cdot 10^{-6} - 1\cdot 10^{-5}$~eV$^2$ \cite{sol98}
inferred from the MSW solution of the solar neutrino problem \cite{msw86} and 
$\D m^2_{\n_\m \n_\t}\simeq (5\cdot 10^{-4}-6\cdot 10^{-3})$~eV$^2$ associated 
with the atmospheric neutrino deficit \cite{atm98}. 
In the following we shall use for numerical
estimates the average of the neutrino masses of the second and third family,
$\bar{m}_\n=(m_{\n_\m}m_{\n_\t})^{1/2} \sim 10^{-2}$~eV.
 
Consider now the simplest class of seesaw models where three right-handed
neutrinos $\n_{R_i}$ are added to the standard model. According to (\ref{yuk})
and the charge assignment in table~1, the Dirac neutrino mass matrix has
the form
\beq\label{dir}
    m_D = h^{(\n)} v =\ \e^a\ \left(\begin{array}{ccc}
    A\e^{d+1} & B\e^{c+1} & C\e^{b+1} \\[1ex]
    D\e^d     & E\e^c     & F\e^b     \\[1ex]
    G\e^d     & H\e^c     & K\e^b
    \end{array}\right)\ v \;,
\eeq
where $A,\ldots, K$ are constants of order one. Correspondingly, the
Majorana mass matrix of the right-handed neutrinos reads,
\beq\label{maj}
    M = h^{(s)} \langle S \rangle = \ \left(\begin{array}{ccc}
    \a\e^{2d} & 0         & 0       \\[1ex]
    0         & \b\e^{2c} & 0       \\[1ex]
    0         & 0         & \g\e^{2b}
    \end{array}\right)\langle S \rangle \;.
\eeq
Here $\a,\ldots,\g$ are constants of order one and, without loss of generality,
we have chosen a basis where $M$ is diagonal and real. Note, that this
diagonalization does not change the hierarchy structure of the Dirac mass
matrix (\ref{dir}). The corresponding
mass eigenstates are the Majorana neutrinos $N_i\simeq \n_{R_i} + \n_{R_i}^c$.
In terms of the Dirac mass matrix (\ref{dir}) and the Majorana mass matrix 
(\ref{maj}) the Majorana mass matrix of the light neutrinos is given by 
\cite{yan79}
\beq\label{seesaw}
m_\n = m_D {1\over M} m_D^T \;.
\eeq
In this expression for the light neutrino masses the dependence on the
chiral charges of the heavy neutrinos drops out and one obtains the
hierarchy structure (\ref{matrix}).

Heavy Majorana neutrinos are likely to play a crucial role for the 
cosmological baryon asymmetry. At high temperatures, above the critical
temperature of the electroweak phase transition, baryon and lepton number
violating processes are in thermal equilibrium \cite{kuz85}. As a
consequence, a primordial lepton asymmetry generated by the out-of-equilibrium
decay of heavy Majorana neutrinos is partially transformed into a
baryon asymmetry \cite{fuk86}. The dominant contribution to the lepton
asymmetry is produced in the decays of $N_1$, the lightest of the
heavy Majorana neutrinos. The corresponding baryon asymmetry is given by,
\beq
Y_B = {n_B\over s} = \k\ C\ {\ve_1\over g_*}\;.
\eeq
Here $\ve_1$ is the CP-asymmetry in the decay of $N_1$, C is the ratio
of lepton and baryon asymmetry, which equals -8/15 in the 
(supersymmetric) standard model with 2 Higgs doublets, 
$g_* \sim 100$ is the number of effectively massless
degrees of freedom and $\k$ is a dilution factor. Its value reflects the 
effect of the various lepton number conserving and violating processes in 
the plasma. In order to determine $\k$ reliably one has to solve the full
Boltzmann equations. To obtain a large asymmetry, the number
density of the heavy neutrinos at high temperatures has to be large enough, 
they have to fall out of equilibrium at $T \sim M_1$, and the $\D L =1$
and $\D L=2$ washout processes have to be sufficiently suppressed. Typical
values of the dilution factor are then $\k \sim 10^{-1}-10^{-2}$,
which corresponds to a baryon asymmetry
\beq\label{basym}
Y_B \sim (10^{-3}-10^{-4})\ \ve_1\;.
\eeq

The CP asymmetry $\ve_1$ can be computed in terms of the neutrino mass 
matrices. For degenerate heavy neutrinos, i.e. $M \propto 1$, $\ve_1$ 
vanishes.
We therefore assume hierarchical heavy neutrino masses, with $d \geq 1$.
In this case $M_1 \ll M_{2,3}$, and one obtains \cite{cov96,buc98},  
\beq
\ve_1={3\over16\pi v^2}\;{1\over\left(m_D^{\dag}m_D\right)_{11}}
         \sum\limits_{i=2,3}
         \mbox{Im}\left[\left(m_D^{\dag}m_D\right)_{1i}^2\right]\
         {M_1\over M_i}\label{cpas}\;.
\eeq
Note, that the CP asymmetry depends on $m_D$ only through $m_D^{\dag}m_D$.
Hence, $\ve_1$ is not directly related to the $\n_\m$-$\n_\t$ mixing angle
in the leptonic charged current. 

From eqs.~(\ref{dir}), (\ref{maj}), (\ref{seesaw}) and (\ref{cpas}) one easily
obtains the dependence of the CP asymmetry and the light neutrino masses
on the hierarchy parameter $\e$,
\bea
\ve_1 &\sim& 10^{-1}\ \e^{2(a+d)}\;,\label{e1}\\
\bar{m}_\n &\sim& \e^{2(a+d)}\ {v^2\over M_1}\;.
\eea 
Using $\bar{m}_\n \sim 10^{-2}$~eV, the corresponding mass of $N_1$,
the lightest of the heavy neutrinos, is given by
\beq
M_1 \sim \e^{2(a+d)}\ 10^{15} \mbox{GeV}\;.
\eeq
Note, that the values of $\ve_1$ and $M_1$ do not depend on the charges
$b$ and $c$ of the heavy neutrinos $N_3$ and $N_2$, respectively.

From eqs.~(\ref{basym}) and (\ref{e1}) one obtains for the baryon asymmetry, 
\beq
Y_B \sim (10^{-4}-10^{-5})\ \e^{2(a+d)}\;.
\eeq
Hence, the baryon asymmetry is quantized in powers of the hierarchy parameter 
$\e^2$. Its magnitude is fixed by the neutrino charges. 

Consider first the simplest case of hierarchical neutrino 
masses with Yukawa couplings of the
third family $\co{(1)}$, i.e. $a=b=0$, $c=1$, $d=2$. One then has 
\beq
M_1 \simeq 10^{10}\ \mbox{GeV}\;,\quad M_3 \simeq 10^{15}\ \mbox{GeV}\;;
\quad Y_B \sim 10^{-9}-10^{-10}\;.
\eeq
These are precisely the parameters chosen in \cite{buc96}. Solving the
Boltzmann equations indeed yields an asymmetry $Y_B \sim 10^{-10}$, 
the baryogenesis temperature is $T_B \sim M_1 \sim 10^{10}$~GeV and
$B-L$ is broken at the GUT scale. 

The same results are obtained for $c=0$. 
Alternatively, one may have smaller Yukawa couplings together with 
a smaller mass hierarchy. This corresponds to $a=0$, $b=c=1$, $d=2$ or
$a=1$, $b=c=0$, $d=1$. In both cases baryogenesis temperature and baryon
asymmetry remain the same, but $M_2 \sim M_3 \sim 10^{12}$~GeV. Hence,
the scale of $B-L$ breaking may be below the GUT scale.

In all cases considered so far the CP asymmetry $\ve_1 \sim 10^{-6}$.
One may think that the charge asignement $a=b=c=0$, $d=1$, which yields
a CP asymmetry $\ve_1 \sim 10^{-4}$, may lead to a larger baryon asymmetry.
However, one then has $M_1 \sim 10^{12}$~GeV. The dependence of the
baryon asymmetry on $\tilde{m}_1 = (m_D^{\dag}m_D)_{11}/M_1$ and $M_1$
has been studied in the non-supersymmetric and in the supersymmetric case
by solving the full Boltzmann equations \cite{plu97}. It turns out that
the dependence on $M_1$ is model dependent. For the above choice of
parameters, with $\tilde{m}_1 \sim \bar{m}_\n\sim 10^{-2}$~eV, the washout 
processes in the supersymmetric case are too strong, and the generated 
asymmetry lies far below the observed value $Y_B \sim 10^{-10}$.
Hence, the observed baryon asymmetry corresponds to the maximal asymmetry.

The baryogenesis temperature $T_B \sim 10^{10}$~GeV is naturally obtained in
hybrid models of inflation. For supersymmetric theories such a large reheating
temperature imposes stringent constraints on the mass spectrum of 
superparticles, in particular on the gravitino mass. Consistency with
primordial nucleosynthesis requires either a very light gravitino,
$m_{\tilde{G}} < 1$~keV \cite{pag82}, a heavy gravitino,  
$m_{\tilde{G}} > 2$~TeV \cite{mor95}, or a gravitino in the mass range
$m_{\tilde{G}} = (10-100)$~GeV as LSP \cite{bol98}, with a higgsino type 
neutralino as next-to-lightest superparticle. In the last case gravitinos
could be the dominant component of dark matter.
 
There are other mechanisms of baryogenesis. In particular in supersymmetric
theories a very large baryon asymmetry may be produced by the coherent
oscillation of scalar fields carrying baryon and lepton number \cite{aff84}.
The generated asymmetry then depends on the initial configuration of scalar
fields. This is in contrast to leptogenesis, as described above. Here,
the baryon asymmetry is entirely determined by neutrino masses and mixings
which, at least in principle, can be measured in precision experiments
probing the leptonic charged current.

Crucial tests of leptogenesis are the Majorana nature of neutrinos
and CP violation in the leptonic charged current. Note, however, that
the expected electron neutrino mass $m_{\n_e} \sim 10^{-5}$~eV is rather small.
Its detection in neutrino-less $\b\b$-decay is an experimental challenge.

Leptogenesis requires that the neutrino mass $\tilde{m}_1 \sim \bar{m}_\n 
= (m_{\n_\m}m_{\n_\t})^{1/2}$ is very
small in order to satisfy the out-of-equilibrium condition for the
decaying heavy neutrino. According to present indications for neutrino 
oscillations, which yield $\bar{m}_\n \sim 10^{-2}$~eV, this is indeed the
case. For the simplest hierarchy pattern one obtains a baryon asymmetry 
$Y_B \sim 10^{-10}$, in agreement with observation.

\end{document}